\documentclass[twocolumn,showpacs,preprintnumbers,amsmath,amssymb,
superscriptaddress,floatfix]{revtex4}
\usepackage{graphicx}
\usepackage{amsmath}
\usepackage{bm}
\usepackage{mathrsfs}
\usepackage{color}
\usepackage{slashed}
\usepackage{dcolumn}
\usepackage[normalem]{ulem}

\newcommand{\be}{\begin{equation}}
\newcommand{\ee}{\end{equation}}
\newcommand{\ba}{\begin{eqnarray}}
\newcommand{\ea}{\end{eqnarray}}

\newcommand{\pr}{\prime}

\begin{document}

\title{The proton-neutron resonance states by solving Schrodinger equation}

\author{Bao-Xi Sun}
\email{sunbx@bjut.edu.cn}
\affiliation{School of Physics and Optoelectronic Engineering, Beijing University of Technology, Beijing 100124, China}

\author{Qin-Qin Cao}
\email{s202166092@emails.bjut.edu.cn}
\affiliation{School of Physics and Optoelectronic Engineering, Beijing University of Technology, Beijing 100124, China}

\author{Ying-Tai Sun}
\email{3071450876@qq.com}
\affiliation{School of Mechanical and Materials Engineering, North China University of Technology, Beijing 100144, China}

\date{\today}

\begin{abstract}
The proton-neutron interaction is investigated by solving the Schrodinger equation, where a Yukawa type of potential with one pion exchanging between the proton and the neutron is assumed. Since the deutron is the unique bound state of the proton-neutron system, the coupling constant is fixed according to the binding energy of the deutron. The scattering process of the proton and the neutron is studied when the outgoing wave condition is taken into account, and two proton-neutron resonance states are obtained by solving the Schrodinger equation, which lie at $1905-i13$MeV and $2150-i342$MeV on the complex energy plane, respectively. It is no doubt that the calculation results would give some hints on the experimental research on the proton-neutron interaction in future.
\end{abstract}


\maketitle

\section{Introduction}

The non-Hermitian physics has caused more interests of theorists in the past decade, and its methods are applied in different areas of physics, such as condensed matter physics, optics and nuclear physics. More detailed statements on this topic can be found in the review articles\cite{Ashida:2020dkc,Meden:2023yng}. However, in the study of exotic hadrons, the non-Hermitian character is not discussed yet.

Two kinds of situations are often discussed in the non-Hermitian physics. Firstly, the Hamiltonian of the system is non-Hermitian, and the eigenfunctions and their biorthogonal normalization are evaluated. In order to study the non-Hermitian properties, a non-Hermitian system is constructed in the laboratory\cite{zhaoerhai,fanyingying}.
In the traditional quantum mechanics, the Hamiltonian of a physical system is defined as a Hermite operator. However, this definition    is based on the constraint of bound states, where the wave function tends to zero at the infinity. If this constraint condition is eliminated, for example, in the scattering process, the whole system will be non-Hermitian although the Hamiltonian of the system still takes a Hermite form in some special representations.
Since the non-Hermitian properties of an operator depend not only on the operator itself but also on the properties of the wave functions, the Hamiltonian in a non-Hermitian system can have complex eigenvalues\cite{Moiseyev}.
It is apparent that the wave function is not zero at the infinity in the scattering process, thus the scattering process can be treated as a non-Hermitian problem. In Ref.~\cite{Moiseyev}, the scattering process in a square well is studied, and the bound states below the threshold and the resonance states above the threshold are obtained respectively by solving the Schrodinger equation with different boundary conditions of the wave function.

The deutron can be regarded as a proton-neutron($pn$) bound state, and the structure of the deutron is investigated by solving the Schrodinger equation in the S-wave approximation, where a Yukawa type of potential is assumed between the proton and the neutron\cite{Zhangyongde}. With the method of variable substitution, the radial Schrodinger equation takes a form of the Bessel equation in the S-wave approximation, and the solution for the bound state corresponds to the Bessel function.
When the scattering process of the proton and the neutron is taken into account, the first and second kinds of Hankel functions are also solutions of the radial Schrodinger equation, and represent the incoming wave and the outgoing wave respectively.
Therefore, the coupling constant of the proton and the neutron in the Yukawa type of potential is determined according to the binding energy of the deutron. Under the outgoing wave condition, the zero point of the second kind of Hankel function would correspond to the proton-neutron resonance state in the scattering process, and the position of the resonance state on the complex energy plane is related to the order of the second kind of Hankel function, which is complex when the outgoing wave condition is satisfied.

In this work, the proton-neutron scattering process is studied by solving the Schrodinger equation when the wave function does not disappear at the infinity, and two $pn$ resonance states are obtained, which lie at $1905-i13$MeV and $2150-i342$MeV on the complex energy plane, respectively. Moreover, this method is extended to study the meson-meson interaction, and  a series of hadronic resonance states are generated dynamically when the meson-meson scattering process is considered\cite{sunbxKKstar}.

The whole article is organized as follows: The $pn$ bound state is discussed shortly in Section~\ref{Sect:deutron}, and then the $pn$  interaction is studied by solving the Schrodinger equation under the outgoing wave condition in Section~
\ref{Sect:pn}. Finally, a summary is given in the last section.

\section{The proton-neutron bound state in a Yukawa potential}
\label{Sect:deutron}

The one-pion-exchanging-potential(OPEP) plays a dominant role in the long range interaction of the proton and the neutron, and
the potential takes a Yukawa form, i.e.,
\be
\label{eq:202307071816}
V(r)=-g^2\frac{e^{-mr}}{d},
\ee
with $m$ the mass of the pion and $g$ the coupling constant.
In Eq.~(\ref{eq:202307071816}), the distance $r$ in the denominator of the potential has been replaced with the range of force $d=\hbar/mc$ approximately.

If the radial wave function $R(r)=\frac{u(r)}{r}$, thus the radial Schrodinger equation with $l=0$ can be written as
\be
\label{eq:202307081218}
-\frac{\hbar^2}{2\mu} \frac{d^2 u(r)}{dr^2}+V(r)u(r)=Eu(r),
\ee
with $\mu=M/2$ the reduced mass, and $M$ the mass of the nucleon.

By using the variable substitution
\be
r \rightarrow x=\alpha e^{-\beta r},~~~~0 \le x \le \alpha,
\ee
with
\be
\alpha=\frac{2g}{\hbar}\sqrt{2 \mu d},~~~~\beta=\frac{1}{2d},
\ee
and
\be
u(r)=J(x),
\ee
the radial Schrodinger equation becomes the $\rho$th order Bessel equation, i.e.,
\be
\label{eq:202307081903}
\frac{d^2 J(x)}{d x^2}+\frac{1}{x} \frac{d J(x)}{d x} +\left[1-\frac{\rho^2}{x^2}\right] J(x)=0,
\ee
with
\be
\rho^2=-\frac{4d^2 2\mu E}{\hbar^2},~~~~E<0.
\ee
It is no doubt that the solution of Eq.~(\ref{eq:202307081903}) is just the $\rho$th Bessel function $J_\rho(x)$\cite{Zhangyongde}.

The deutron can be regarded as a proton-neutron bound state, and its root-mean-squared radius is less than 2fm. Therefore,
when $r \rightarrow +\infty$, the radial wave function $R(r) \rightarrow 0$, which implies that $u(r)=J_\rho(\alpha e^{-\beta r})=J_\rho(0)$ with $\rho \ge 0$. Moreover, when $r \rightarrow 0$, $u(r) \rightarrow 0$, and it means
\be
\label{eq:boundcondition}
J_\rho(\alpha)=0.
\ee
Therefore, as a bound state of the proton-neutron system, the deutron corresponds to the first nonzero zero-point of the Bessel function actually.

\begin{figure}
\includegraphics[width=0.5\textwidth]{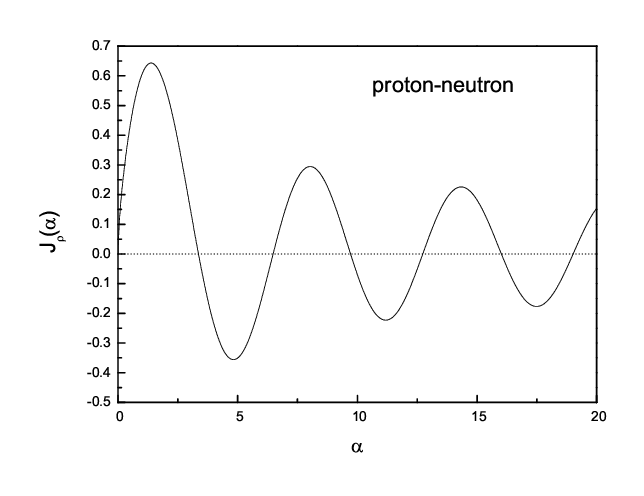}
\caption{
The Bessel function $J_\rho(\alpha)$ with $\rho=0.6547$ for the proton-neutron system with the first nonzero zero-point at $\alpha=3.3622$, which is assumed to correspond to the deutron.
}
\label{fig:zeropointpn}
\end{figure}

The binding energy of the deutron is about $2.224575\pm0.000009$MeV\cite{deutronexp}, i.e., $E=-2.224575$MeV, so the order of the Bessel function $\rho \sim 0.6547$. Therefore, the first nonzero zero-point of $J_\rho(\alpha)$ lies at $\alpha=3.3622$, and the coupling constant in the Yukawa potential is calculated according to
\be
\label{eq:coupling}
\frac{g^2}{\hbar c}=\frac{\hbar \alpha^2}{8 \mu c d},
\ee
which takes a value of ${g^2}/{\hbar c}=0.4203$.
The Bessel function $J_\rho(\alpha)$ with $\rho = 0.6547$ as a function of $\alpha$ is depicted in Fig.~\ref{fig:zeropointpn}, where the zero-points of the Bessel function are shown explicitly.

Since the second zero-point of $J_0(\alpha)$ is $\alpha=5.520$, for the fixed value of $\alpha$, which is less than 5.520, there is only one determined real value of $\rho$ so as to make $J_\rho(\alpha)=0$. Therefore, in the S-wave approximation, only one bound state of the proton-neutron system is obtained when $\alpha<5.520$.

\section{The proton-neutron scattering process in a Yukawa potential}
\label{Sect:pn}

The Hankel functions $H_\rho^{(1)}(x)$ and $H_\rho^{(2)}(x)$ are also two independent solutions of the Bessel equation\cite{sunbxKKstar}. Actually, $H_\rho^{(1)}(x) e^{-i \omega t}$ represents a wave along the positive direction of the $x$ axis, while $H_\rho^{(2)}(x) e^{-i \omega t}$ corresponds to a wave along the negative direction of the $x$ axis. When the $pn$ scattering process is discussed, the general solution of Eq.~(\ref{eq:202307081903}) is given as
\be
\label{eq:202307081915}
\psi(x)=D H_\rho^{(1)}(x) +D^\pr H_\rho^{(2)}(x).
\ee

In the scattering process, the wave function exists in the infinity. However, the wave function must take a finite value when $r\rightarrow 0$, which implies $\psi(x)$ tends to zero.
Therefore, as discussed in Ref.~\cite{sunbxKKstar}, the bound and virtual states are determined with the incoming wave condition
\be
\label{eq:incoming}
H_\rho^{(1)}(\alpha)=0,
\ee
while the resonance state is generated dynamically according to the outgoing wave condition
\be
\label{eq:outgoing}
H_\rho^{(2)}(\alpha)=0.
\ee
Actually, the incoming wave condition in Eq.~(\ref{eq:incoming}) is equivalent to the condition of bound states in Eq.~(\ref{eq:boundcondition}) since the $J_\rho(\alpha)$ and $H_\rho^{(1)}(\alpha)$ have the same zero-point when their orders $\rho$ are the same as each other.

\begin{figure}
\includegraphics[width=0.5\textwidth]{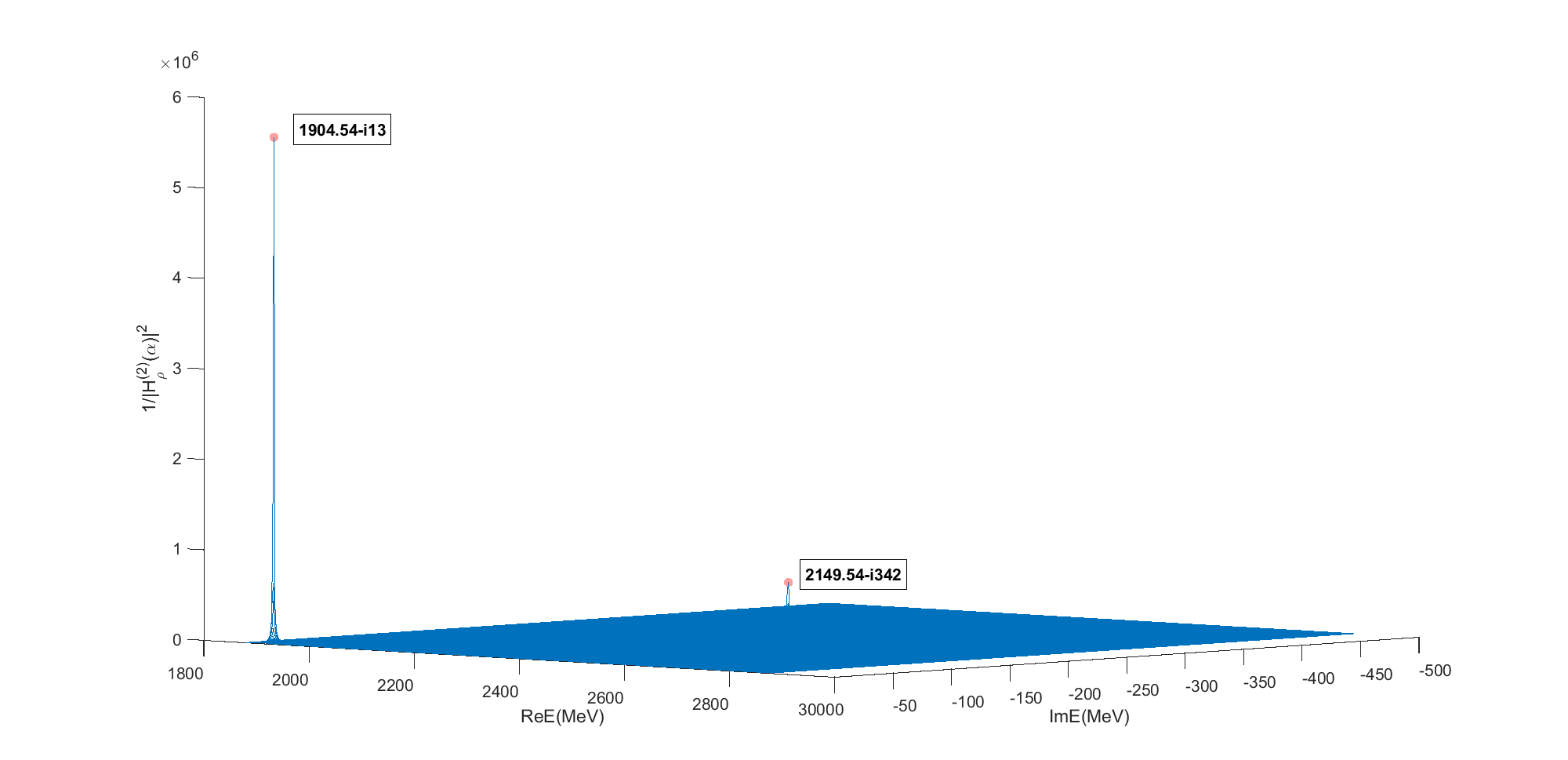}
\caption{$1/|H_\rho^{(2)}(\alpha)|^2$ .vs. the complex energy $E$ with $\alpha=3.3622$ in the proton-neutron case. The pole of $1/|H_\rho^{(2)}(\alpha)|^2$ corresponds to a zero-point of the second kind of Hankel function $H_\rho^{(2)}(\alpha)$, which represents a $p n$ resonance state, as labeled in the figure.}
\label{fig:pn}
\end{figure}

With the outgoing wave condition in Eq.~(\ref{eq:outgoing}) and  $\alpha=3.3622$, two proton-neutron resonance states can be obtained at $1905-i13$MeV and $2150-i342$MeV, respectively.
Here the value of $1/|H_\rho^{(2)}(\alpha)|^2$ is calculated when the total energy of the system changes, and thus the zero-point of the Hankel function $H_\rho^{(2)}(\alpha)$ corresponds to the pole of $1/|H_\rho^{(2)}(\alpha)|^2$ on the complex energy plane respectively, as shown in Fig.~\ref{fig:pn}.
Although the experimental data on the $pn$ resonance state are insufficient, it is no doubt that our calculation would give some hints on the experimental search on these resonance states.

\section{Summary}

In this work, the proton-neutron system is studied by solving the Schrodinger equation under different boundary conditions. By fitting the coupling constant in the one-pion-exchange potential with the binding energy of the deutron, two resonance states are generated dynamically when the outgoing wave condition is taken into account. It is more important that the calculation results indicate that there are intrinsic relations between the bound state and resonance states of the interaction system.



\end{document}